\documentclass[%
 reprint,
superscriptaddress,
 amsmath,amssymb,
aps,
prl,
]{revtex4-2}
\usepackage{chemformula}
\usepackage{graphicx}
\usepackage{dcolumn}
\usepackage{bm}

\begin{document}

\preprint{APS/123-QED}

\title{Spin Current Density Functional Theory of the Quantum Spin-Hall Phase}

\author{William P. Comaskey}
\affiliation{Department of Physics, College of Arts and Science, Florida State University, Tallahassee FL, 32310, USA.}

\author{Filippo Bodo}
\affiliation{Dipartimento di Chimica, Università di Torino, Via Giuria 5, 10125 Torino, Italy}
\affiliation{Department of Chemistry, Southern Methodist University, Dallas, Texas, USA}

\author{Alessandro Erba}
\email{alessandro.erba@unito.it}
\affiliation{Dipartimento di Chimica, Università di Torino, Via Giuria 5, 10125 Torino, Italy}

\author{Jose L. Mendoza-Cortes}
\email{jmendoza@msu.edu}
\affiliation{Department of Physics, College of Arts and Science, Florida State University, Tallahassee FL, 32310, USA.}
\affiliation{Department of Chemical Engineering \& Materials Science, Michigan State University, East Lansing, Michigan 48824, United States.}

\author{Jacques K. Desmarais}
\email{jacqueskontak.desmarais@unito.it}
\affiliation{Dipartimento di Chimica, Università di Torino, Via Giuria 5, 10125 Torino, Italy}

\date{\today}

\begin{abstract}
The spin current density functional theory (SCDFT) is the generalization of the standard DFT to treat a fermionic system embedded in the effective external field produced by the spin-orbit coupling interaction. Even in the absence of a spin polarization, the SCDFT requires the electron-electron potential to depend on the spin currents $\mathbf{J}^x$, $\mathbf{J}^y$ and $\mathbf{J}^z$, which only recently was made possible for practical relativistic quantum-mechanical simulations [Phys. Rev. B {\bf 102}, 235118 (2020)]. Here, we apply the SCDFT to the quantum spin-Hall phase and show how it improves (even qualitatively) the description of its electronic structure relative to the DFT. We study the Bi (001) 2D bilayer and its band insulator to topological insulator phase transition (via $s+p_z \leftrightarrow p_x +ip_y$ band inversion) as a function of mechanical strain. We show that the explicit account of spin currents in the electron-electron potential of the SCDFT is key to the appearance of a Dirac cone at the $\Gamma$ point in the valence band structure at the onset of the topological phase transition. Finally, the valence band structure of this system is rationalized using a simple first-order $\mathbf{k} \cdot \mathbf{p}$ quasi-degenerate perturbation theory model.
\end{abstract}

\maketitle


\section{Introduction}


Quantum spin-Hall (QSH) systems (or topological insulators, TIs) are insulating in the bulk while they exhibit gapless edge states (ES) that hold spin currents with little or no dissipation. Notably, and in contrast to its relative (i.e. the quantum Hall phase), appearance of the conducting ES does not require breaking of time-reversal symmetry (TRS)~\cite{bernevig2006quantum,moore2010birth,kane2006new}. Existence of the QSH phase was first proposed in the analytical model for graphene of Kane and Mele (KM) in 2005 from an inclusion of spin-orbit coupling (SOC) in the effective tight-binding Hamiltonian of Haldane~\cite{canna_applesI,haldane1988model}. The same authors later showed that the QSH phase is in fact topologically distinct from other insulating states, being characterized by a $\mathbb{Z}_2$ topological invariant~\cite{canna_applesII}. This implies that the conducting ES with desirable spin-transport properties are, in fact, particularly robust against perturbations. 

Despite these early successes, the KM model is limited by the lack of renormalization of the SOC potential by the electron-electron ($ee$) interaction. At about the same time, Bencheikh showed how the spin current density functional theory (SCDFT) of Vignale and Rasolt~\cite{w1} would provide a formulation of the DFT for a fermionic system in the effective external field produced by SOC~\cite{w2}. It follows that the SCDFT provides the necessary theoretical framework to the description of the missing $ee$ terms on the QSH phase. On the other hand, a conventional DFT treatment would not guarantee an accurate description because its $ee$ term does not satisfy certain invariance relations in the presence of SOC~\cite{desmaraistrsb}. For TRS preserving states (such as those relevant to the QSH effect), the SCDFT dictates that the $ee$ potential must depend on the electron density $n$, and on the three spin current densities $\mathbf{J}^x$, $\mathbf{J}^y$ and $\mathbf{J}^z$~\cite{w12,desmarais2020adiabatic,bodo2022spin}. The latter are exactly those physical quantities at the core of the peculiar transport phenomena in the QSH phase, and whose effect on the $ee$ potential has been missing in previous treatments~\cite{hirahara2006role,koroteev2004strong,Yakovkin2019,xia2009observation,Hsieh2008,zhang2009topological,murakami2006quantum,wada2011localized,chege2020origin}, with a notable exception being the study by Trushin and G{\"o}rling of the AlBi and SnTe 3D TIs by use of the exact-exchange (EXX) functional~\cite{w12}. With the latter approach, however, electron correlation is neglected. Some of the authors of this Letter have recently formulated a practical strategy to SCDFT calculations including the effects of electron correlation~\cite{desmarais2020adiabatic,bodo2022spin,desmarais2019fundamental}. 

In this Letter, we apply the SCDFT to the QSH phase and show how the corresponding electronic structure differs even qualitatively from that predicted by the standard DFT. As a prototypical example, we study the band insulator to topological insulator phase transition in the Bi (001) 2D bilayer~\cite{hirahara2006role,koroteev2004strong,yang2012spatial,murakami2006quantum,drozdov2014one,wada2011localized}. We demonstrate that modification of the $ee$ potential by the SOC-induced spin currents in SCDFT not only captures the expected band inversion phenomenon at the phase transition but also predicts the existence of previously overlooked Dirac fermions (DFs) emerging as quasiparticles close to the top of the valence band at the $\Gamma$ point at the onset of the topological phase transition. 

\begin{widetext}
\begin{center}
\begin{figure}[h!!]
\includegraphics[width=0.93\linewidth]{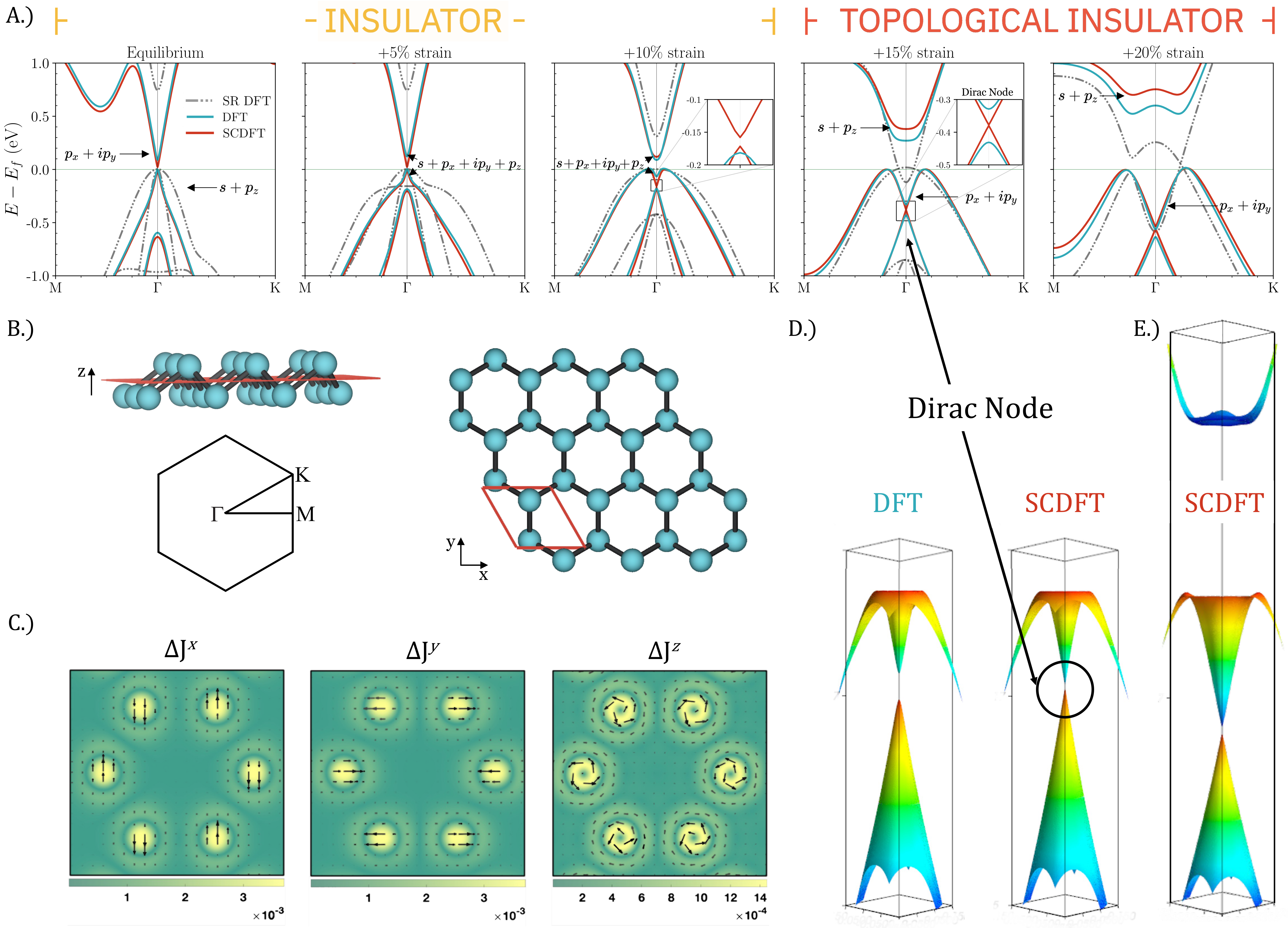}
\caption{Electronic structure of the Bi (001) bilayer as described by the DFT and SCDFT. A) Band structure along the $M-\Gamma-K$ path (shown to the right of the structure in B) as a function of strain (from the equilibrium lattice up to a +20\% strain). Dashed black lines are from scalar-relativistic DFT (without SOC), blue continuous lines are from DFT with SOC (that is, from the potential of Eq. \ref{eqn:dft}), and red continuous lines are from SCDFT with SOC (that is, from the potential of Eq. \ref{eqn:scdft}). The ``character'' of the top of valence and bottom of conduction bands in the vicinity of the $\Gamma$ point is reported (in terms of the leading atomic orbital contributions). B) Atomic structure of the Bi (001) bilayer: side view (top) and top view (bottom). On the top, the plane between the two atomic layers is shown as a red plane. On the right the $M-\Gamma-K$ path is shown for the $R\overline{3}m$ space-group.  C) Spin current densities ${\bf J}^x$, ${\bf J}^y$ and ${\bf J}^z$ for the +15\% strain case in the plane highlighted in the top-left panel in B). The color identifies the absolute value in the selected plane while the length and direction of the superimposed black arrows represent the magnitude and direction of their in-plane Cartesian components. D) 3D representation of the top valence bands at +15\% strain in the vicinity of $\Gamma$ as obtained with the DFT (left) and SCDFT (right). E) 3D representation of the top valence bands and bottom conduction band in the vicinity of $\Gamma$ at +20\% strain as obtained with the SCDFT.}
\label{fig:bande}
\end{figure}
\end{center}
\end{widetext}

All calculations are performed with a developmental version of the \textsc{Crystal17} program~\cite{Dovesi2018,MPP2017,desmarais2019spinI,desmaraistrsb,desmarais2021spin2}. The computational details are reported in the electronic supporting information (ESI)~\cite{ESI_QSH} (see also Refs. ~\cite{doll2001analytical,doll2001implementation,doll2006analytical,civalleri2001hartree,metz2000small,heifets2015ab,lebedev1976quadratures,lebedev1977spherical,towler1996density,perdew1996generalized,seidl1996generalized,gorling1994exact,perdew1996rationale,adamo1999toward,heyd2003hybrid,dick2012advanced,voon2009kp,kirtman1981simultaneous} included therein). Our DFT and SCDFT calculations are based on hybrid exchange-correlation (xc) functionals, employing the generalized gradient approximation (GGA). Indeed, in the GGA of the SCDFT for TRS preserving electronic states, an adiabatic connection formula for the xc potential $\hat{v}_{xc}^\text{SCDFT}$ may be derived, by exploiting the short-range behaviour of the exchange hole, yielding~\cite{w3,desmarais2020adiabatic,bodo2022spin}:
\begin{eqnarray}
\label{eqn:scdft}
\hat{v}_{xc}^\text{SCDFT} \left[ n,\boldsymbol{J}^x,\boldsymbol{J}^y,\boldsymbol{J}^z \right] =  \hat{v}_{c} \left[ n\right] + (1 - \alpha) \ \hat{v}_{x} \left[ n\right] \nonumber \\
+ \alpha \ \hat{X}^\text{SCDFT} [n,\boldsymbol{J}^x,\boldsymbol{J}^y,\boldsymbol{J}^z] \; ,
\end{eqnarray}
where $\alpha$ is the dimensionless fraction of exact exchange (with its associated potential $\hat{X}^\text{SCDFT}$), and where $\hat{v}_{x}$ and $\hat{v}_{c}$ are exchange and correlation potentials in the GGA. On the other hand, in the DFT, the corresponding expression reads:
\begin{equation}
\label{eqn:dft}
\hat{v}_{xc}^\text{DFT} \left[ n \right] =  \hat{v}_{c} \left[ n\right] +
 (1 - \alpha) \ \hat{v}_{x} \left[ n\right] + \alpha \ \hat{X}^\text{DFT} [n] \; .
\end{equation}
The EXX potential $\hat{X}^\text{DFT}$ in Eq. (\ref{eqn:dft}) may be obtained from a unitary transformation of $\hat{X}^\text{SCDFT}$, using an approach described in Ref. \cite{bodo2022spin}. In Eq. (\ref{eqn:scdft}), $\boldsymbol{J}^c \left( \mathbf{r} \right)$ with $c=x,y,z$ are spin current densities defined as follows (for simplicity we report the expression for the one-electron non-periodic case): 
\begin{equation}
\label{eq:scd}
\boldsymbol{J}^c \left( \mathbf{r} \right) =\frac{1}{2i} \left\{ \psi^\dagger \left( \mathbf{r} \right) \boldsymbol{\sigma}^c \left[ \boldsymbol{\nabla} \psi \left( \mathbf{r} \right)\right] - \left[ \boldsymbol{\nabla} \psi^\dagger \left( \mathbf{r} \right)\right] \boldsymbol{\sigma}^c \psi \left( \mathbf{r} \right) \right\} \; ,
\end{equation}
where $\psi \left( \mathbf{r} \right)$ is a one-electron two-component (2c) spinor, $\boldsymbol{\sigma}^c$ is the $c$-th Pauli matrix and $i=\sqrt{-1}$. Therefore, a comparison of predictions from Eqs. (\ref{eqn:scdft}) and (\ref{eqn:dft}) provides a probe to study the effect on the electronic structure of including the spin current densities in the $ee$ potential. Here, the xc GGA potential is PBE and $\alpha=$10\%.

Starting from the conventional rhombohedral $R\overline{3}m$ bulk crystal structure of Bi~\cite{wei2019volatile}, we extract a 2D layered structure by cutting over a plane orthogonal to the [001] direction, leading to a buckled-honeycomb bilayer shown in Figure \ref{fig:bande}B. Then, we fully relax the structure (in terms of both lattice parameters and atomic positions) at the scalar-relativistic (SR) level, under the only constraints provided by the hexagonal layer group symmetry. Strained configurations are explored by constraining the lattice parameters to larger values compared to the equilibrium ones (+5, +10, +15 and +20\%) and by letting the atoms relax within the strained cells. 

The evolution of the electronic band structure as a function of strain is reported in Figure \ref{fig:bande}A as predicted by scalar relativistic (SR) calculations (i.e. without SOC) and by inclusion of SOC within a DFT and SCDFT framework. For the Bi (001) bilayer, the bands are constrained both by TRS [$\epsilon^\uparrow(\mathbf{k})=\epsilon^\downarrow(-\mathbf{k})$] and space-inversion symmetry [ $\epsilon^\uparrow(\mathbf{k})=\epsilon^\uparrow(-\mathbf{k})$] leading to $\epsilon^\uparrow(\mathbf{k})=\epsilon^\downarrow(\mathbf{k})$, such that the bands are doubly-degenerate and SOC-induced spin-splitting does not occur. The DFT bands resulting from calculations with $\alpha =$ 0 are also provided in Fig. S1 of the ESI. 


At the equilibrium structure (Figure \ref{fig:bande}A, leftmost panel), the effect of SOC is manifested through the drastic reduction of the direct band gap at $\Gamma$ (from 0.75 eV to 0.02 eV), as already noted from previous DFT calculations~\cite{Yakovkin2019}. Moving to the right panels, as the lattice parameter is expanded, a typical band-inversion (BI) profile is observed. The BI can be tracked through an analysis of the principal atomic-orbital components of the crystalline single-particle orbitals at $\Gamma$. At the equilibrium configuration, the top of the valence band is of mixed $s+p_z$ character, while the bottom of the conduction band is of the type $p_x + i p_y$. As the lattice is expanded, these components are initially mixed in both states (see +5\% and +10\% strained configurations) and eventually inverted at +15\% strain, a typical signature of the topological phase transition (TPT)~\cite{maciejko2011quantum,bernevig2006quantum,chege2020origin}. We also note that, at the TPT, the band gap opens significantly.  

Remarkably, at the onset of the TPT (i.e. at +15\% strain), the SCDFT predicts the formation of a doubly-degenerate Dirac cone (DC) at the top of the valence band at $\Gamma$, which is not predicted by the DFT (see Figure \ref{fig:bande}A and \ref{fig:bande}D). That is, the formation of such a distinctive electronic feature is only predicted by inclusion of the spin current densities in the $ee$ potential. Even more remarkably, this same DC was noted in previous angle-resolved photoemission spectroscopy experiments, but was attributed as being due to the interaction with the substrate~\cite{yang2012spatial}. 


\begin{figure}[b!]
\includegraphics[width=8.5cm]{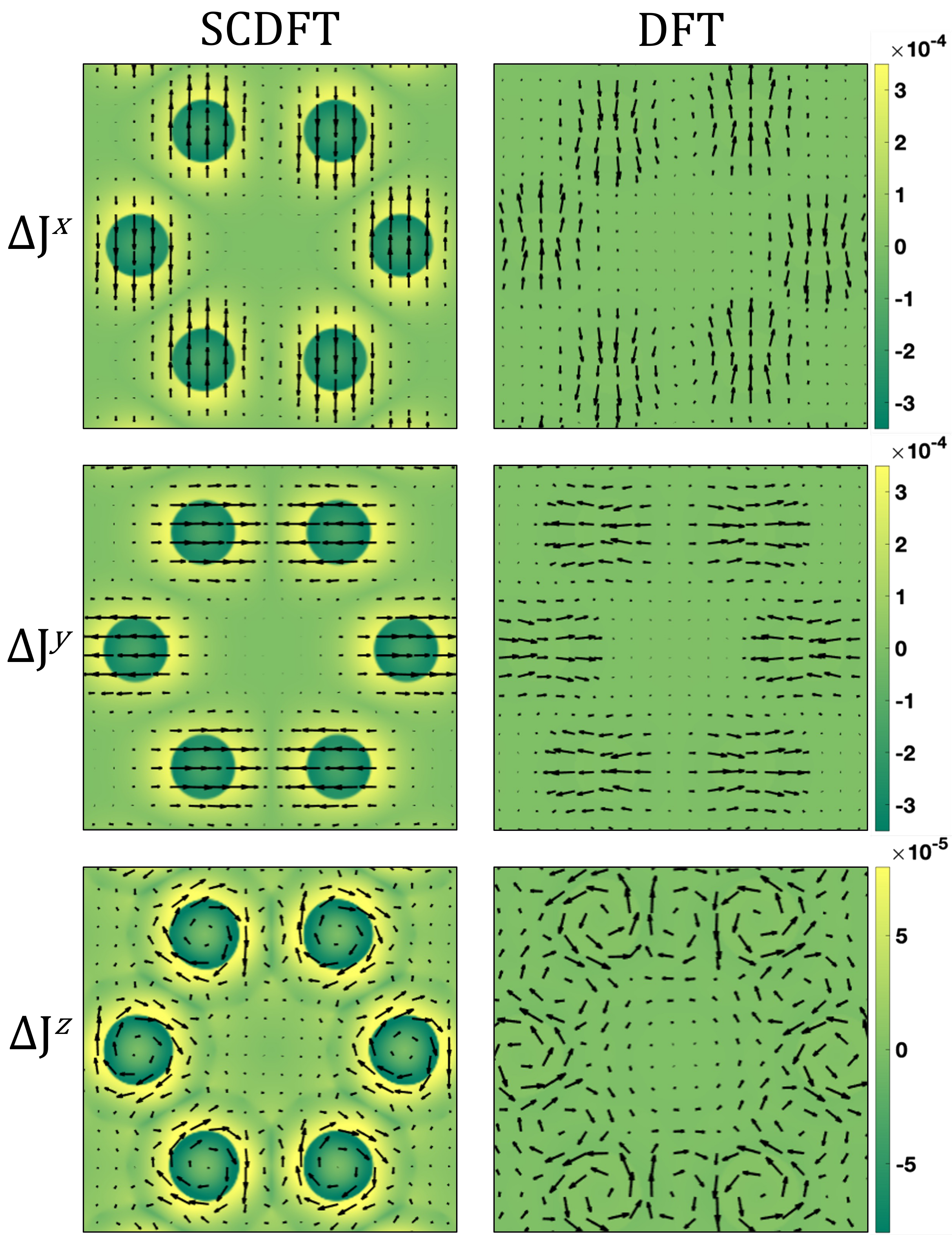}
\caption{Effect of orbital-relaxation through the self-consistent field process on spin current densities in DFT and SCDFT calculations. The reported quantities are differences between the final electronic solution and the initial one: $\Delta {\bf J}^c = {\bf J}^c_\textup{final} - {\bf J}^c_\textup{initial}$, with $c=x,y,z$. The panels show color maps of the spatial distribution of these quantities in the same plane as in Figure \ref{fig:bande}C.}
\label{fig:jspin}
\end{figure}

We rationalize the formation of this DC by means of a first-order $\mathbf{k} \cdot \mathbf{p}$ quasi-degenerate perturbation theory model of the valence band structure, inspired by Kane's model~\cite{kane1957band,orlita2014observation}. Formal aspects are given in the ESI. We show that when the SOC-induced gap at the $\Gamma$ point, $E_\Gamma$, is large, the model predicts a dispersion relation for the four highest-lying valence bands exactly along a doubly-degenerate DC:
\begin{equation}
\epsilon \left( \mathbf{k}\right) = \pm v \vert \mathbf{k} \vert \; ,
\end{equation}
where $v$ is the band velocity. Thus, the formation of the DC by the SCDFT correlates with the larger SOC-induced gap of $E_\Gamma = 0.40$ eV compared to that of $E_\Gamma = 0.28$ eV by the DFT where the DC does not emerge. As discussed above, this is entirely due to the dependence on the three spin current densities $\boldsymbol{J}^x,\boldsymbol{J}^y$ and $\boldsymbol{J}^z$ of the SCDFT $ee$ potential of Eq. (\ref{eqn:scdft}). Figure \ref{fig:bande}C shows the spatial distribution of these vector quantities in the central plane of the Bi (001) bilayer.

The leading effect of including the three spin current densities in the $ee$ potential has been discussed above and is further highlighted and quantified below. Let us stress that, in the presence of SOC, spin current densities can be computed from Eq. (\ref{eq:scd}) both within a DFT and SCDFT approach. However, we show in Figure \ref{fig:jspin} that their inclusion in the $ee$ potential in the SCDFT case results in their significant evolution along the self-consistent field process (i.e. orbital relaxation), which, on the other hand, is lacking in the DFT. Figure \ref{fig:jspin} shows color maps of the orbital relaxation of the three spin current densities along the SCF: $\Delta {\bf J}^c = {\bf J}^c_\textup{final} - {\bf J}^c_\textup{initial}$, with $c=x,y,z$, where the initial quantities are calculated from a second-variational treatment of SOC (i.e. after a single diagonalization of the $2c$ Hamiltonian built from SR orbitals). The color scale describes the magnitude of the quantities while the arrows represent their local orientation and magnitude in the selected plane (same as in Figure \ref{fig:bande}). The figure clearly shows that DFT calculations are incapable of accounting for the orbital-relaxation of the spin currents. On the other hand, the SCDFT calculation allows for the significant build-up of spin currents along the SCF process. Ultimately, this enhanced role of the spin currents in the SCDFT compared to the DFT is at the core of the formation of the DC in the valence bands of the Bi (001) bilayer at the onset of the TPT.

\paragraph{Acknowledgments}
This work was supported in part through computational resources and services provided by the Institute for Cyber-Enabled Research at Michigan State University. J.K.D. is grateful to the National Science and Engineering Research Council of the Government of Canada for a Postdoctoral Fellowship application \# 545643. We are grateful to John S. Tse for a fruitful discussion.


%

\end{document}